\documentclass[a4paper,11pt]{article}
\usepackage{pos}

\title{Parametrisation scale invariance of Parton Distribution Function fits}

\author[a]{Ivan A. Godino}
\author[a,b]{Eva D. Z. Groenendijk}
\author*[c,d]{Tanjona R. Rabemananjara}

\affiliation[a]{Tif Lab, Dipartimento di Fisica, Università degli Studi di Milano, Italy}
\affiliation[b]{INFN, Sezione di Milano, Via Celoria 16, I-20133 Milano, Italy}
\affiliation[c]{Department of Physics and Astronomy, Vrije Universiteit, NL-1081 HV Amsterdam, The Netherlands}
\affiliation[d]{Nikhef Theory Group, Science Park 105, 1098 XG Amsterdam, The Netherlands}

\emailAdd{eva.groenendijk@unimi.it}

\abstract{
In global PDF analyses, parton distribution functions (PDFs) are parametrised at a fixed input scale $Q_0$ and evolved to higher scales using the DGLAP equations. Since QCD evolution is fully determined within perturbation  theory, the fitted PDFs should, in principle, be independent of the arbitrary choice of $Q_0$. In this work, we test this within the NNPDF framework by performing a series of fits at different starting scales, namely $Q_0 \in \{1.0,1.25,1.5,1.65\}$ GeV. We find that the resulting PDFs are consistent within uncertainties and the fit quality remains stable across this range, therefore confirming the invariance of PDF determinations with respect to the choice of starting scale.
}

\FullConference{
}

\begin{document}
\maketitle

\section{Introduction}

%
Contemporary global PDF analyses determine PDFs by parametrising their functional forms at a fixed input scale $Q_0$, typically of order $\mathcal{O}(1)$ GeV, and evolving them to higher scales via the DGLAP evolution equations. The PDFs $f_i(x,Q_0)$ are thus fitted at this initial scale, while their scale dependence is governed by perturbative QCD (pQCD). In principle, since QCD evolution is unitary, the resulting PDFs should be independent of the arbitrary choice of the starting scale $Q_0$.
The input scale is typically chosen to have a low value, of order $\mathcal{O}(1)$ GeV. The reason  is that small differences that appear at high-$Q$ are known to be larger when evolved towards lower energy scales. This is because of asymptotic freedom, which causes PDFs to be more similar at higher energy scales. If the PDF is therefore fitted at a higher energy scale, a small difference there will result in a large difference at low energy scale, while fitting the PDF at lower scale reduces this effect because a large difference at low $Q$ becomes smaller at higher scales.
In the NNPDF4.0 methodology~\cite{nnpdf40}, a fixed starting scale of $Q_0=1.65$ GeV is used, which is slightly above the charm threshold of $m_c=1.502$ GeV. The fits should however be independent of this choice. In this note we show that indeed PDFs parametrised at different starting scales in a range of $Q_0 \in \{1.0,1.25,1.5,1.65\}$ GeV agree within uncertainties. The new starting scales that are tested are all below the charm quark threshold, which also makes this a test for evolution across the mass threshold. We also show that fit quality remains similar in this range.

\section{Method}

In order to compare PDF fits at different parametrisation scales, we keep the NNPDF methodology and theory exactly the same, and only vary $Q_0$. For the treatment of quark thresholds we use the values of the masses recommended by the LHC Higgs working group~\cite{higgs}. All other parameters, such as the CKM matrix, and the electromagnetic coupling are also taken from there. Partonic cross sections are computed at NNLO, and for the perturbative evolution we use the exact solutions to the DGLAP equations. As for fitting methodology, the same methods are used as in NNDPF4.0 \cite{nnpdf40}. The PDF parametrisation is

\begin{equation}\label{eq:parametrisation}
    xf_i(x,Q_0;\mathbf{\theta}) = A_i x^{\alpha_i^{\text{eff}}}(1-x)^{\beta_i^{\text{eff}}}\text{NN}_i(x;\mathbf{\theta}),
\end{equation}

\noindent with $i$ denoting the PDF flavour. The preprocessing exponents $\alpha_i^{\text{eff}}$ and $\beta_i^{\text{eff}}$ are varied in a range determined by iteration of the fit as explained in \cite{nnpdf30}. The PDFs are fitted in the evolution basis as in \cite{nnpdf40} and QCD evolution is performed using \textsc{eko}\cite{eko}. A subset of the total NNPDF4.0 dataset is used for the PDF fitting, excluding the $t\bar{t}$ and jet datasets because of computational resources. For the neural network architecture we use the outcome of the NNPDF4.0 hyperoptimisation as explained in \cite{nnpdf40}. This means that the architecture is optimised for a starting scale of $Q_0=1.65$ GeV, while here we use it at different scales.

\section{Results}

We compare PDFs fitted at different parametrisation scales $Q_0 \in \{1.0,1.25,1.5,1.65\}$ GeV. Firstly we compare the fit qualities corresponding to these different values of $Q_0$, after which we compare the resulting PDFs themselves. We compare the fits at a low energy, namely $Q=1.65$ GeV, since the low-energy regime is where differences are most likely to appear. We also looked at the fits at a higher scale of $100$ GeV, at which indeed all differences disappear as expected, so we do not show them here. Finally, we also look at the behaviour of the small- and large-$x$ asymptotic exponents $\alpha_{a}$ and $\beta_a$, respectively.

\subsection{Fit quality for different starting scales}

A first check of the independence on $Q_0$  is to compare the quality of the agreement of the fit with the data as a function of $Q_0$, as measured by the experimental $\chi^2$ (see Ref.~\cite{nnpdf40} for the definition of the experimental $\chi^2$). Independence requires the $\chi^2$ to be independent of the particular value of $Q_0$,   which is indeed the case, looking at Table~\ref{tab:chi2}. We also looked at the training lengths (TL) and training and validation losses as defined in \cite{nnpdf40}, which all agree within uncertainties. Therefore we conclude that the general quality of the fit does not change varying over these parametrisation scales.

\begin{table}[h!]
    \centering
    \begin{tabular}{c|c|c}
        $Q_0$ (GeV) & $\chi^2$ & <TL>\\
        \hline
        1.0 & 1.15122 & 10700±3200\\
        1.25 & 1.14711 & 9900±3600 \\
        1.5 & 1.14753 & 10000±3700 \\
        1.65 & 1.14707 & 10300±3400 \\
    \end{tabular}
    \caption{\textit{The $\chi^2$'s and the average training lengths of the fits with different starting scales. The TL (the number of epochs before the optimal stopping point) is averaged over MC replicas.}}
    \label{tab:chi2}
\end{table}

\noindent From Table. \ref{tab:chi2} it can be seen that at $Q_0=1.0$ GeV, the $\chi^2$ is slightly higher than the others. This might be due to the fact that the fitting metholodogy has been hyperoptimised for a starting scale of $Q_0=1.65$ GeV. At $Q_0=1.0$ GeV the fit could benefit from a slightly different architecture that allows for more freedom, because of the stronger impact of QCD evolution in this regime. This being said, the fit quality is still almost as good as before, which shows the robustness of the methodology. Furthermore, training length and training and validation losses are compatible within uncertainties between $Q_0=1.0$ and $Q_0=1.65$ GeV, so the value of the $\chi^2$ could also  be a statistical fluctuation.

\subsection{PDF comparison}

\begin{figure}[h!]
    \centering
\includegraphics[width=0.45\textwidth]{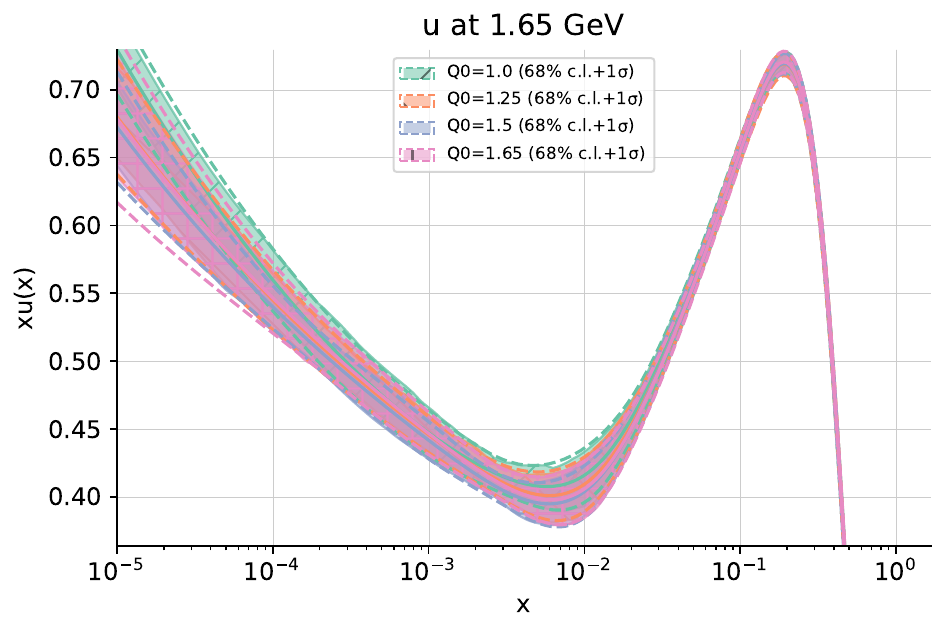}
\includegraphics[width=0.45\textwidth]{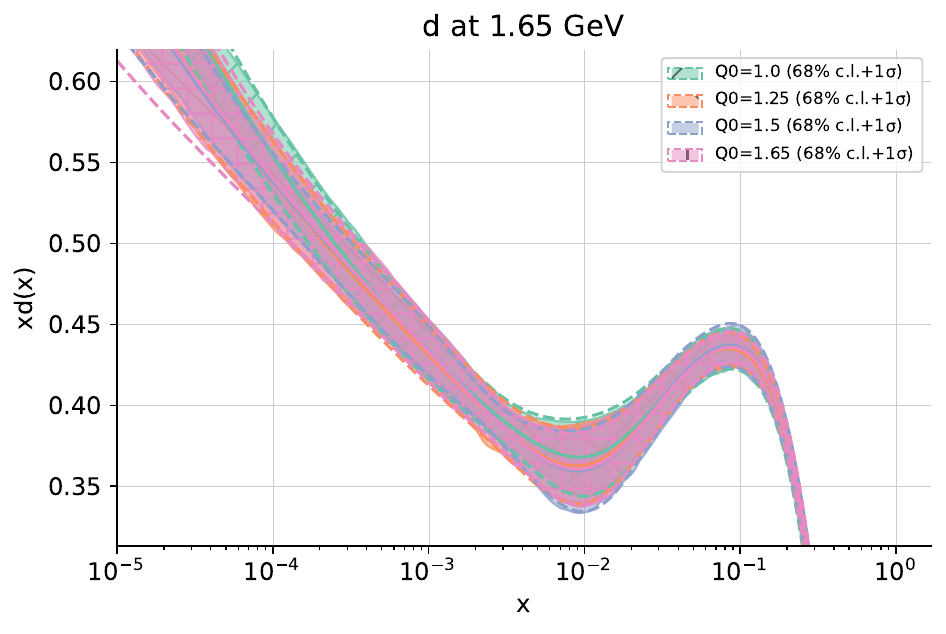}
\includegraphics[width=0.45\textwidth]{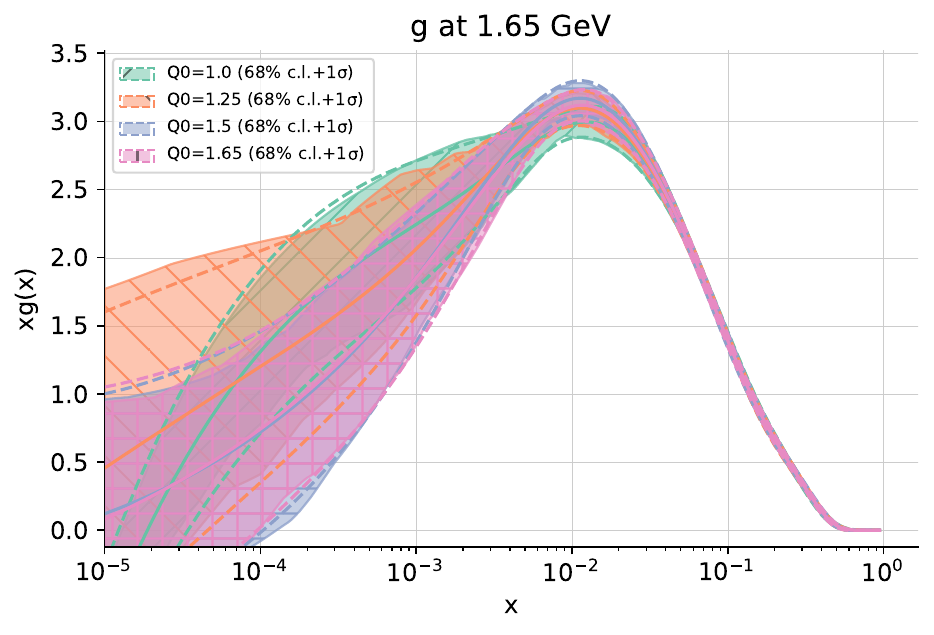}
\includegraphics[width=0.45\textwidth]{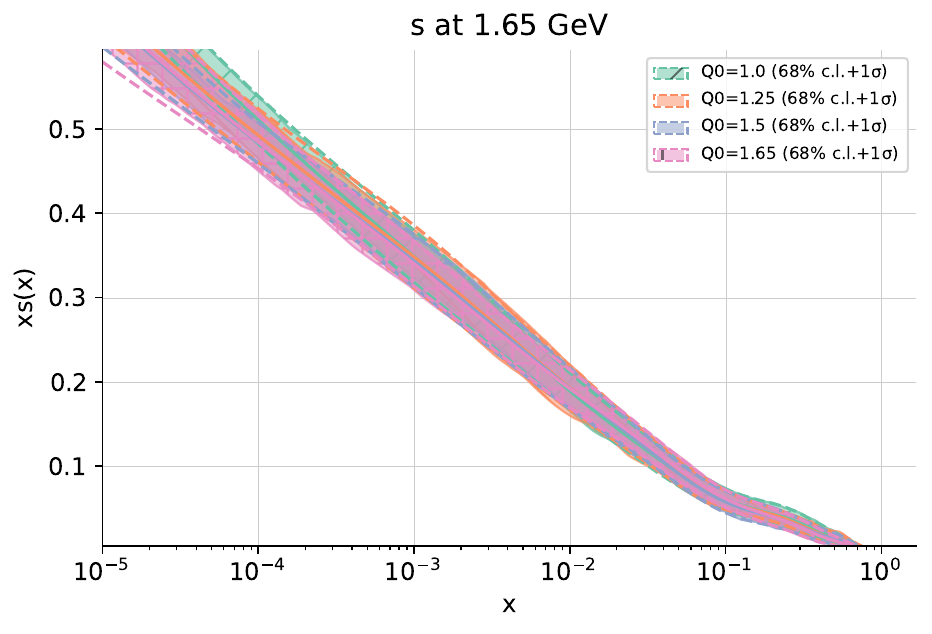}
    \caption{\textit{The effect of choosing a different parametrisation scale for fitting the up, down and strange quark PDF and the gluon PDF. The comparison is between parametrisation scales of $1.0, 1.25, 1.5$ and $1.65$ GeV. All PDFs are shown at the scale $Q=1.65$ GeV.}}
    \label{fig:pdfs165}
\end{figure} 

    \label{fig:ratio}

The PDFs fitted at different scales agree within uncertainties for all flavours. Fig. \ref{fig:pdfs165} shows the gluon and the up, down and strange quark PDFs at $Q=1.65$ GeV, and the same can be seen for the other flavours. Since we compare PDFs fitted below the charm threshold with a PDF fit above the charm threshold, this shows that matching across the thresholds has no effect on the results. Furthermore, since we fit the charm quark PDF, this shows the results for the intrinsic charm PDF in \cite{ic_2022} are stable with respect to evolution below the charm threshold. As mentioned before, the differences at low scale are expected to be larger, which can be seen in Fig. \ref{fig:pdfs165} in the gluon PDF. At higher energy scales we checked that this difference indeed disappears.
\\
\begin{figure}[hbp!]
    \centering
\includegraphics[width=0.45\textwidth]{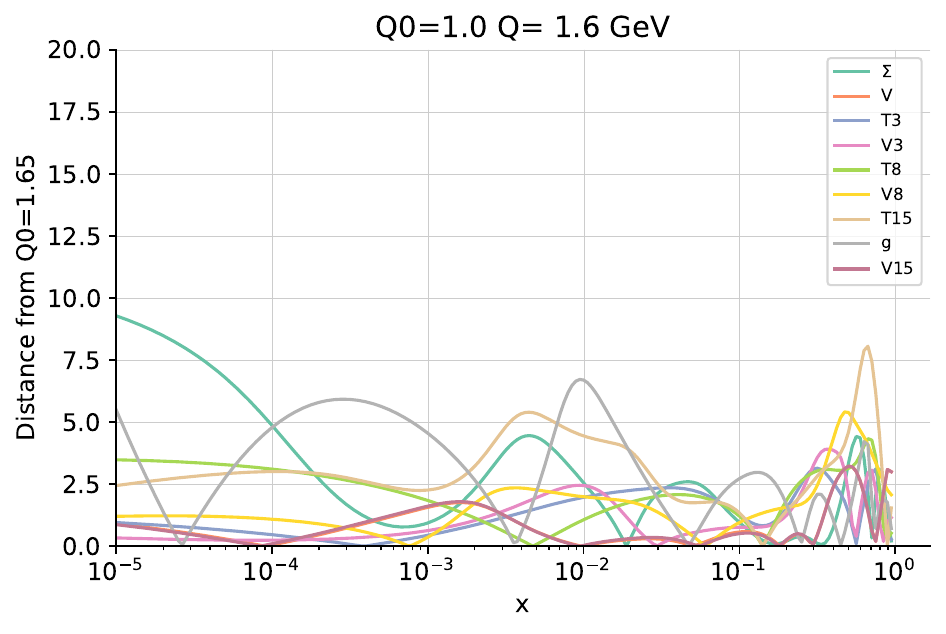}
\includegraphics[width=0.45\textwidth]{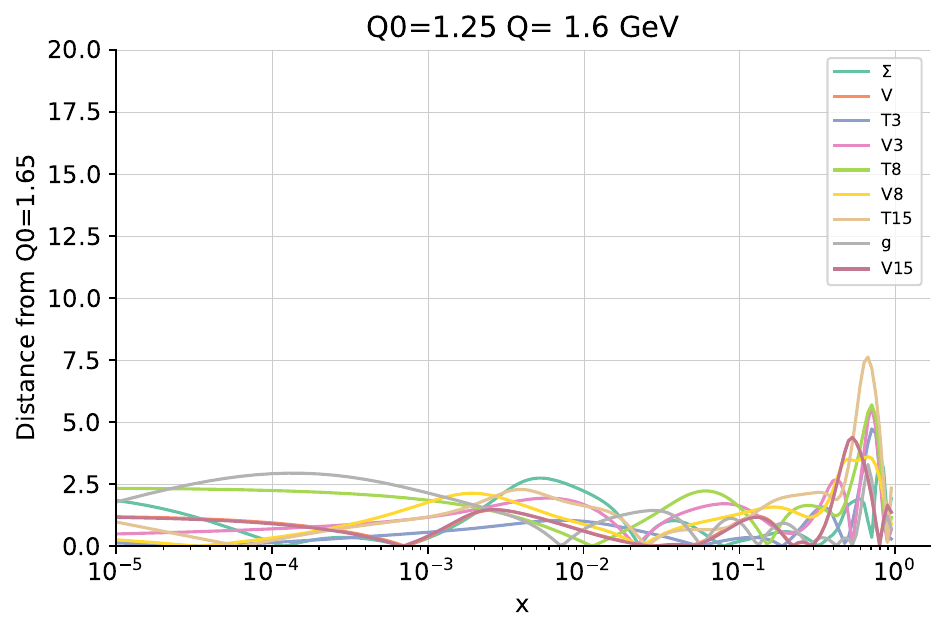}
    \caption{\textit{The distance of the central values of the evolution basis PDFs with $Q_0=1.0$ (left) and $Q_0=1.25$ (right) from the ones with $Q_0=1.65$ GeV.}}
    \label{fig:distances}
\end{figure}


\noindent Fig. \ref{fig:distances} shows the distances between fits with $Q_0=1.0$ ($Q_0=1.25$) and $Q_0=1.65$ GeV. The definition of the distance can be found in \cite{ball_pdfs}: $d=10$ corresponds to a $1 \sigma$ deviation, while $d=1$ corresponds to complete statistical equivalence, i.e. replicas extracted from the same probability distribution. For $Q_0=1.25$, PDFs are almost equivalent, and for $Q_0=1.0$ the difference is up to $0.5 \sigma$ in some regions. The central values of these PDFs are within a difference of $1 \sigma$ everywhere.

\subsection{Asymptotic exponents}

The asymptotic behaviour of PDFs for small-$x$ (or large-$x$, respectively) is described  by the asymptotic exponents $\alpha_i$ (or $\beta_i$, respectively)~\cite{asy_exps} defined as

\begin{align}
    xf_i(x,Q_0) & \xrightarrow{x\rightarrow 0} x^{\alpha_i}, \nonumber \\
    xf_i(x,Q_0) & \xrightarrow{x\rightarrow 1} (1-x)^{\beta_i}.
\end{align}

\begin{figure}[h!]
    \centering
\includegraphics[width=0.45\textwidth]{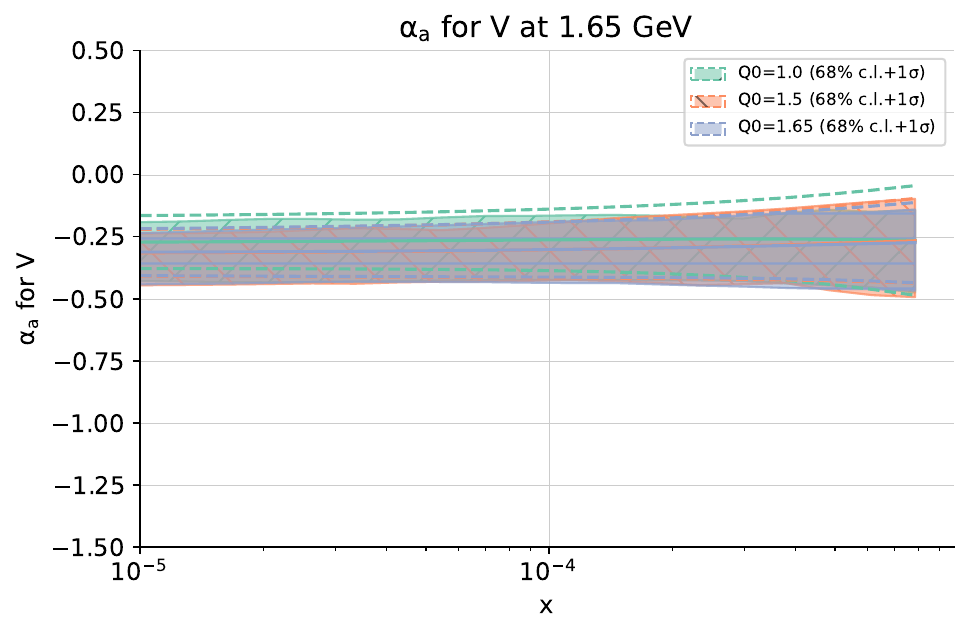}
\includegraphics[width=0.45\textwidth]{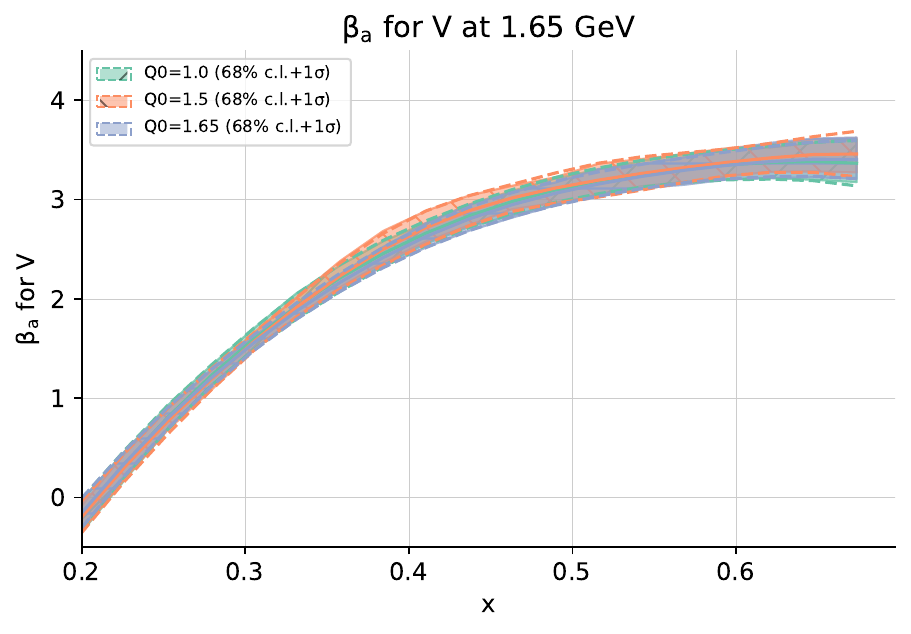}
\includegraphics[width=0.45\textwidth]{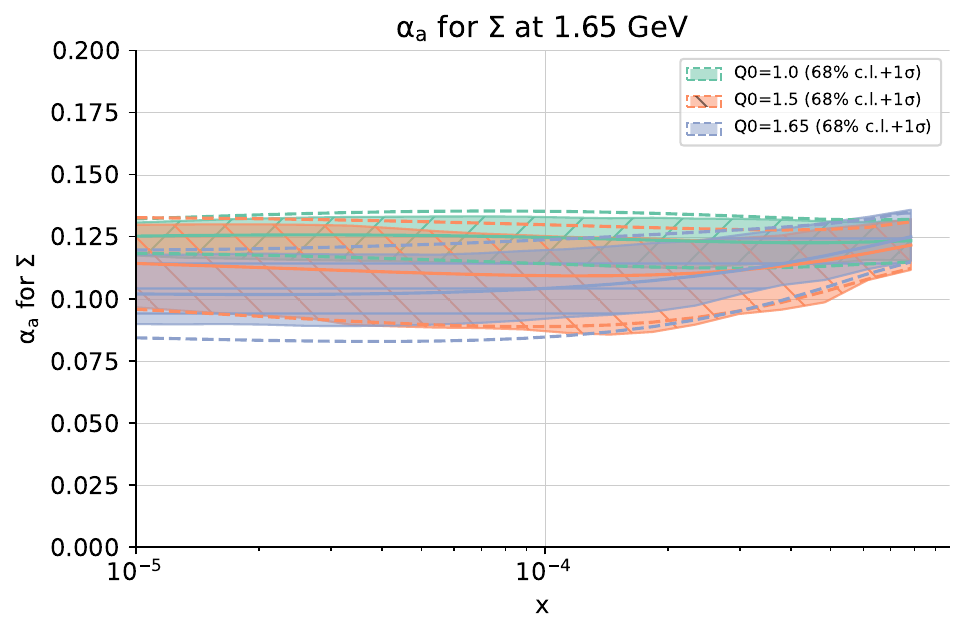}
\includegraphics[width=0.45\textwidth]{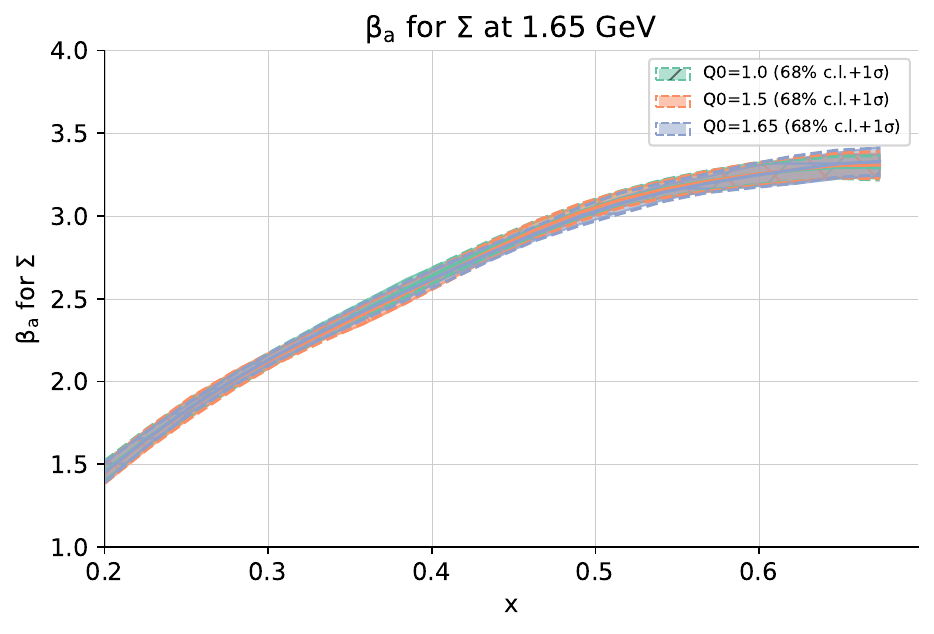}
    \caption{\textit{The small- and large-$x$ asymptotic exponents for the $V$ and $\Sigma$ PDFs fitted at different starting scales $Q_0 \in \{1.0,1.5,1.65\}$. The exponents are computed at the scale $Q=1.65$ GeV.}}
    \label{fig:asy_q}
\end{figure}

\noindent The exponents $\alpha_i$ and $\beta_i$ therefore contain the information on the asymptotic behaviour of the PDF extrapolated outside the data region. The values of $\alpha_i$ and $\beta_i$ can be determined as

\begin{equation}
    \alpha_i(x,Q) = \frac{\partial \ln(xf_i(x,Q))}{\partial \ln x}, \ \ \
    \beta_i(x, Q) = \frac{\partial \ln(xf_i(x,Q))}{\partial \ln (1 - x)},
\end{equation}

\noindent and are thus computed differently than the effective exponents in Eq. \ref{eq:parametrisation}. Fig. \ref{fig:asy_q} shows the asymptotic exponents for the valence PDF $V = (u-\bar{u})+(d-\bar{d})+(s-\bar{s})$ and the singlet $\Sigma = (u+\bar{u})+(d+\bar{d})+(s+\bar{s})+(c+\bar{c})$, all computed at the same scale $Q=1.65$ GeV but using PDFs parametrised at different scales. For these PDFs the exponents agree within uncertainties for all starting scales. We checked that this is also true for the other flavours. This means that the behaviour of the PDFs when  extrapolated outside the data region also does not depend on the parametrisation scale. 

\section{Conclusion}

In conclusion, we find that indeed the NNPDF methodology is starting scale invariant in the range we tested, even when moving across the charm threshold.  Even though the neural network was hyperoptimised for the particular choice of $Q_0=1.65$ GeV, its performance remains stable upon changes in $Q_0$.  We thus showed the flexibility of the method and the robustness of the implementation of scale evolution.

\bibliographystyle{JHEP}
\bibliography{references}

\end{document}